\documentclass[preprint]{aastex631}

\shorttitle{Locating the CSM Emission within N103B}
\graphicspath{{./}{figures/}}

\begin{document}

\title{Locating the CSM Emission within the Type Ia Supernova Remnant N103B}

\correspondingauthor{Benson Guest}
\email{bguest1@umd.edu}

\author[0000-0003-4078-0251]{Benson T. Guest}
\affiliation{Department of Astronomy, University of Maryland, College Park, MD 20742, USA}
\affiliation{NASA Goddard Spaceflight Center, Greenbelt, MD 20771, USA}
\affiliation{Center for Research and Exploration in Space Science and Technology, NASA/GSFC, Greenbelt, MD 20771, USA}

\author[0000-0003-2379-6518]{William P. Blair}
\affiliation{The William H. Miller III Department of Physics and Astronomy, Johns Hopkins University, 3400 N. Charles Street, Baltimore, MD 21218, USA}

\author[0000-0002-2614-1106]{Kazimierz J. Borkowski}
\affiliation{North Carolina State University, Raleigh, NC 27607, USA}

\author[0000-0002-9886-0839]{Parviz Ghavamian}
\affiliation{ Department of Physics, Astronomy and Geosciences, Towson University, Towson, MD 21252, USA}

\author{Sean P. Hendrick}
\affiliation{Millersville University, Millersville, PA 17551, USA}

\author[0000-0002-4134-864X]{Knox S. Long}
\affiliation{Space Telescope Science Institute, 3700 San Martin Drive, Baltimore MD 21218, USA}
\affiliation{Eureka Scientific, Inc. 2452 Delmer Street, Suite 1, Oakland, CA 94602-3017}

\author[0000-0003-3850-2041]{Robert Petre}
\affiliation{NASA Goddard Spaceflight Center, Greenbelt, MD 20771, USA}

\author[0000-0002-7868-1622]{John C. Raymond}
\affiliation{Harvard-Smithsonian Center for Astrophysics, Cambridge, MA 02138, USA}

\author[0000-0002-4410-5387]{Armin Rest}
\affiliation{Space Telescope Science Institute, 3700 San Martin Drive, Baltimore, MD 21218, USA}

\author[0000-0001-8858-1943]{Ravi Sankrit}
\affiliation{Space Telescope Science Institute, 3700 San Martin Drive, Baltimore, MD 21218, USA}

\author[0000-0002-5044-2988]{Ivo R. Seitenzahl}
\affiliation{University of New South Wales, Australian Defence Force Academy, Canberra, ACT 2600, Australia}

\author[0000-0003-2063-381X]{Brian J. Williams}
\affiliation{NASA Goddard Spaceflight Center, Greenbelt, MD 20771, USA}




\begin{abstract}
We present results from deep \textit{Chandra} observations of the young Type Ia supernova remnant (SNR) 0509-68.7, also known as N103B, located in the Large Magellanic cloud (LMC). The remnant displays an asymmetry in brightness, with the western hemisphere appearing significantly brighter than the eastern half. Previous multi-wavelength observations have attributed the difference to a density gradient and suggested circumstellar material origins, drawing similarities to Kepler's SNR. We apply a clustering technique combined with traditional imaging analysis to spatially locate various emission components within the remnant. We find that O and Mg emission is strongest along the blast wave, and coincides with \textit{Spitzer} observations of dust emission and optical emission from the non-radiative shocks. The abundances of O and Mg in these regions are enhanced relative to the average LMC abundances and appear as a distinct spatial distribution compared to the ejecta products, supporting the circumstellar medium (CSM) interpretation. We also find that the spatial distribution of Cr is identical to that of Fe in the interior of the remnant, and does not coincide at all with the O and Mg emission.

\end{abstract}



\section{Introduction} \label{sec:intro}
Type Ia supernovae are the result of a thermonuclear detonation of a white dwarf star at the Chandrasekhar mass. The path to explosion is likely to have multiple origins. The simple picture is either a single degenerate system where a white dwarf accumulates mass from a non-degenerate companion star, or the merger of two white dwarfs in a double degenerate system. Each of these paths has numerous theoretical sub-types and possibilities (e.g \cite{Maoz2014}). Proving which origin a particular remnant has, and determining the dominant Type Ia channel remains an active area of investigation. One approach is to look for evidence of interaction with a circumstellar medium (CSM) remaining from mass outflow from the binary system prior to the explosion. This might be expected from a single degenerate progenitor system, but not from a double degenerate. Another method is through analysis of ejecta abundances and structure. 

Here we present results from a deep \textit{Chandra} observation of the supernova remnant (SNR) 0509--68.7, often labelled as N103B, located in the Large Magellanic Cloud (LMC). 
Spitzer observations (\cite{Williams2014}) suggest that the remnant is interacting with a dense CSM (n$_0\sim 10$~cm$^{-3}$), drawing similarities to \textit{Kepler}'s SNR which shows comparable densities (\cite{Blair2007,Williams2012}). \cite{Li2021} used Multi-Unit Spectroscopic Explorer (MUSE) observations obtained with the Very Large Telescope (VLT) to study the dense knots of CSM finding densities $\geq 10,000 $~cm$^{-3}$. Currently, these are the only two known Type Ia remnants which show dense CSM interaction hundreds of years after the explosion.

N103B spans 30$''$ in diameter, which corresponds to 7.2 pc at the LMC distance of 50~kpc. \cite{Hughes95} used X-ray observations with \textit{ASCA} to identify N103B as the remnant of a Type Ia supernova, overturning a previous assumption of a core-collapse progenitor by \cite{Chu88} based on the proximity to an H II region and the OB association NGC~1850. The Type Ia identifier has been supported by numerous further studies (\cite{Lewis03,Lopez11,Yang13,Yamaguchi2014}). The star formation history surrounding N103B was studied by \cite{Badenes09}, finding vigorous star formation in the recent past suggesting the progenitor or its companion may have been relatively massive.
A light echo from the SN was found by \cite{Rest05}, from which they derived an age of $\sim 860$ years. Further support for this young age is provided by expansion measurements from \textit{Chandra} observations separated by 17 years \citep{Williams2018}, where the overall average shock velocity is found to be in excess of 4000 km s$^{-1}$, and measurements of balmer dominated shocks with average velocity of $2070\pm60$~~km~s$^{-1}$ by \cite{Ghavamian2017}. The Balmer dominated filaments are in locations with significant preshock neutral hydrogen. The difference between the Balmer line velocity and measured X-ray proper motion velocity is consistent with measurements of other SNRs e.g. Tycho with Balmer line velocities of $\sim 2000-3000$km s$^{-1}$ and X-ray proper motion velocities of $\sim 4000-6000$km s$^{-1}$ (\cite{Ghavamian2001,Williams2017}).

The X-ray spectrum of N103B, shown in Figure~\ref{fig:FullRemnantSpectrum}, shows strong lines from the ``standard'' elements typically seen in Type Ia SNR spectra (Si, S, Ar, Ca, Fe). Lines from O and Mg are seen as well. Whether the O and Mg results primarily from the CSM or from unburnt ejecta remains an open question. \cite{Burkey2013} developed an X-ray analysis technique utilizing a data clustering algorithm to identify the location of components characteristic of CSM and ejecta within a remnant. They applied their technique to Kepler's SNR, locating within the remnant the areas with pronounced O and Mg emission and showing that they are spatially coincident with CSM emission, rather than ejecta. With such obvious similarities between Kepler and N103B \cite{Williams2014}, an obvious question to ask is whether the O and Mg seen in N103B originates from the CSM or the ejecta.

We apply a modified version of this technique to the N103B observations, combined with traditional imaging and spectroscopic analysis. We show that the most significant O and Mg emission appears to follow a distinct spatially separated distribution from the intermediate mass ejecta dominated emission. Additionally, we show that the distribution of the O and Mg emission morphologically correlates with emission from the CSM observed in other wavelengths.

\section{Observations}
N103B was observed early in the \textit{Chandra} mission for $\sim40$~ks (PI: G. Garmire) in 1999. On February 3, 2003 a $\sim30$~ks observation (PI: S. Portegies Zwart) had N103B fall within the field of view; however, the target was a LMC star cluster, and N103B is far off axis where \textit{Chandra's} spatial resolution is substantially lower. As a result, these data are not used for spectral analysis in this work.  A 400~ks program (PI: B. Williams) was split into 12 segments and completed between March 20 and June 1 2017. The details are given in Table \ref{tab:ObsSummary} with all observations using the ACIS-S array, and the 1999 and 2017 observations positioning N103B on \textit{Chandra's} optical axis pointing of the S3 chip.
The data were processed using the \textit{CIAO} version 4.11 \textit{chandra\_repro} script and CalDB version 4.8.3.
For spectral modeling we fit the observations simultaneously, finding strong residuals from the 1999 and 2003 observations compared to the 2017 observations in the soft $\lesssim 1$ keV range. There appears to be a systematic difference between the epochs, where the best fit models over predict the flux in the early observations. A future study may investigate this difference to determine whether there is conclusive evidence for intrinsic brightening of the source, but we consider the most likely source of this discrepancy is the contaminant buildup on the ACIS detectors that primarily affects the softer energies (\cite{Marshall2004}). For the purposes of this paper, we will therefore use only the 2017 observations for spectral fitting which provide the largest sample of self-consistent data and exclude those from the earlier epochs to avoid unnecessarily increasing our systematic errors. We retain the early observations for our imaging analysis.

\begin{table}
    \centering
    \begin{tabular}{c|c|c}
        ObsID & Year & Exp Time (ks) \\\hline
        125*     &   1999*    &   36.7*\\
        3810*    &   2003*    &   29.7*\\
        18018	&	2017	&	39.5	\\
        18019	&	2017	&	59.3	\\
        18020	&	2017	&	27.2	\\
        19921	&	2017	&	16.9	\\
        19922	&	2017	&	41.4	\\
        19923	&	2017	&	58.3	\\
        20042	&	2017	&	19.8	\\
        20053	&	2017	&	11.2	\\
        20058	&	2017	&	43.8	\\
        20067	&	2017	&	29.7	\\
        20074	&	2017	&	31.2	\\
        20085	&	2017	&	14.9	\\\hline
      
    \end{tabular}
    \caption{Breakdown of the \textit{Chandra} observations totalling 459.6 ks. *The 1999 and 2003 observations are not used in the spectral analysis.}
    \label{tab:ObsSummary}
\end{table}


\section{Methods \& Results} 

\subsection{Comparison with Other Young Type Ia SNRs}

\begin{figure}
    \centering
    \includegraphics[width=\columnwidth]{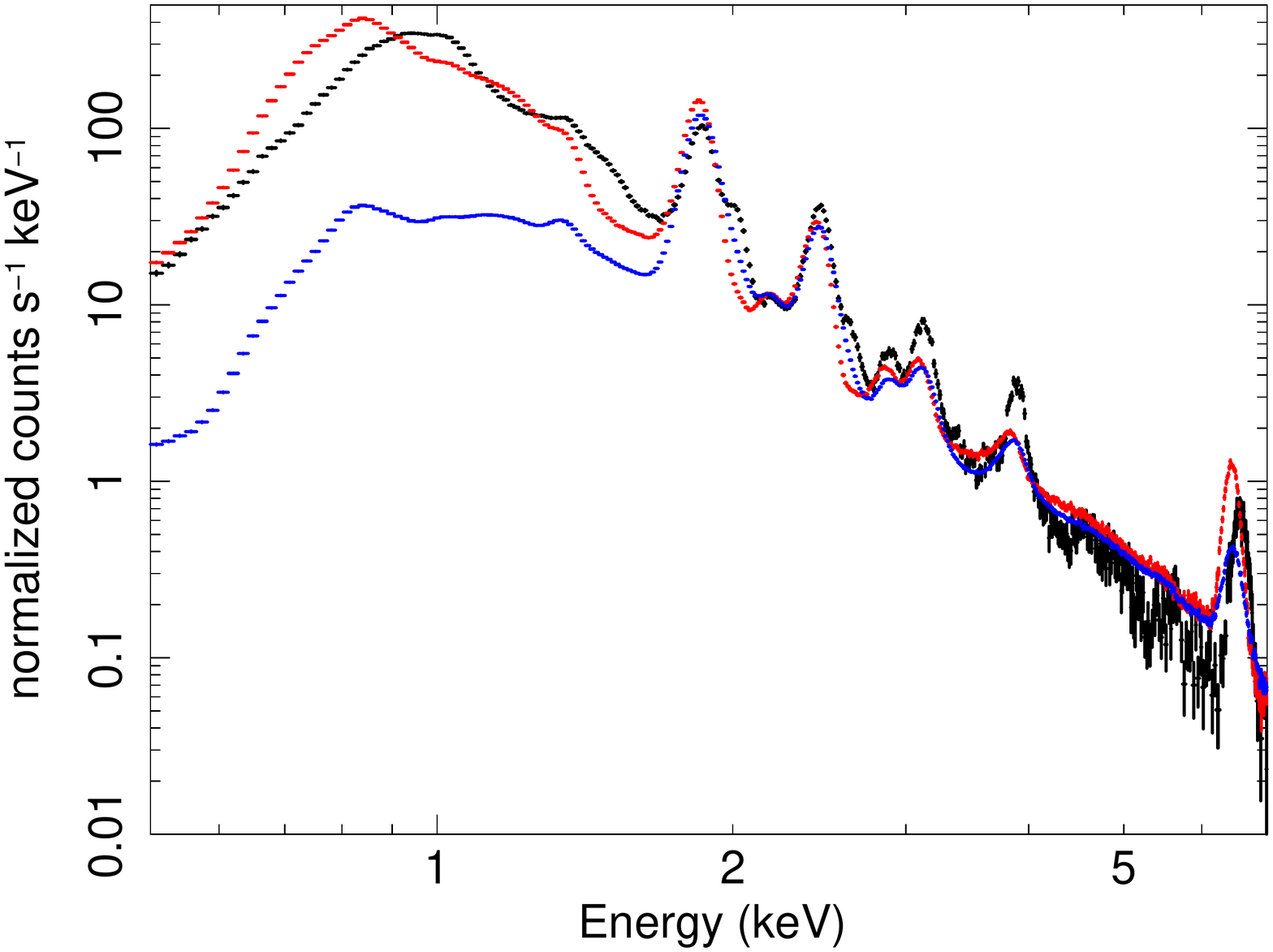}
    \caption{Integrated spectra of N103B (black), Kepler (red), and Tycho (blue). The N103B and Kepler spectra have been scaled up to align the 2.1-2.2 keV continuum with that of the Tycho spectrum.}
    \label{fig:FullRemnantSpectrum}
\end{figure}

We extracted a global spectrum from a circle encompassing the $\sim30''$ diameter remnant, with a background taken from a 27-35$''$ radius annulus surrounding the remnant. The spectrum is plotted in black in Figure \ref{fig:FullRemnantSpectrum} along with \textit{Chandra} spectra of Kepler's SNR and Tycho's SNR in red and blue, respectively. The spectra are arbitrarily scaled to be normalized at 2 keV for display purposes. 

These three remnants are all the result of Type Ia SNe in the last $\sim 1000$ yr, yet their spectra differ substantially. N103B and Kepler show remarkable agreement, particularly compared to Tycho. Kepler and N103B are roughly an order of magnitude brighter at 1 keV than Tycho is, relative to the rest of the spectrum. This is similar to the situation in the mid-IR, where {\it Spitzer} observations of these remnants also show that Kepler and N103B are much more luminous than Tycho (\cite{Blair2007,Williams2013,Williams2014}), despite being at a relatively comparable evolutionary state. The simplest explanation of the major differences in the IR and X-ray spectra from these remnants is that  Kepler's SNR and N103B are expanding into very dense material, {\it two} orders of magnitude higher than Tycho. The subtle differences in Ar and Ca line emission at $\sim 3.1$ and $\sim 3.9$~keV respectively, and the energy centroid of the Fe K$\alpha$ line at $\sim 6.5$~keV are explained by the age of N103B being roughly twice that of Kepler, leading to differences in ionization equilibrium.

\begin{table}
    \centering
    \begin{tabular}{c|c}
        Band & Energy (keV) \\\hline\hline
        O Low & 0.6--0.64 \\
        O Line & 0.64--0.72 \\
        O High & 0.72--0.78 \\
        Fe Peak & 0.73--1.2\\
        Mg Low & 1.2--1.29 \\
        Mg Line & 1.29--1.4 \\
        Mg High & 1.4--1.5\\
        Si Low & 1.6--1.75 \\
        Si Line & 1.75--1.95 \\
        Clustering Si High & 1.95-2.2 \\
        Si Ly$\alpha$ & 1.95--2.1\\
        Si High & 2.1--2.3\\
        S Low & 2.1--2.3 \\
        S Line & 2.3--2.59\\
        S High & 2.59--2.75\\
        Ar Low & 2.94--3.03\\
        Ar Line & 3.03--3.25\\
        Ar High & 3.25--3.4\\
        Ca Low & 3.65--3.8\\
        Ca Line & 3.8--4.0\\
        Ca High & 4.0--4.2\\
        Cr Line & 5.45--5.75\\
        Fe Low & 6.0--6.25\\
        Fe Line & 6.25-6.8\\
        Fe High & 6.8--7.05\\
\hline
    \end{tabular}
    \caption{Band Energies used for Clustering and Equivalent width line images.}
    \label{tab:BandEnergies}
\end{table}

\subsection{Narrow Band Images}
\begin{figure}
    \centering
    \includegraphics[width=\columnwidth]{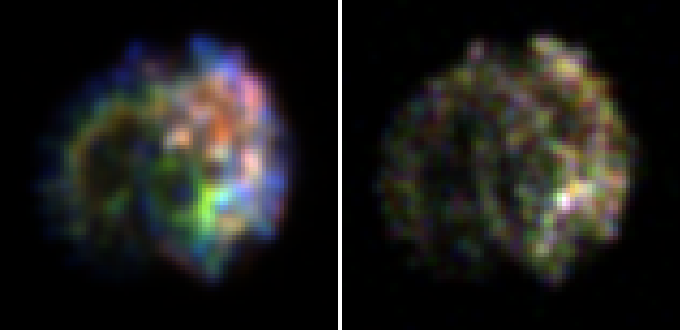}
    \caption{Left: Red: 0.5-0.8 keV (O), Green: 1.2-1.4 keV (Mg), Blue: 1.75-1.95 keV (Si), Right: Red: S, Green: Ar, Blue: Ca. (Band energies listed in Table \ref{tab:BandEnergies}.) The distinct areas of different colors in the left image highlight the different distributions of the emitting material. The image on the right shows no distinct regions where one band dominates over the other two, highlighting the uniformity of the ejecta elements. }
    \label{fig:RGB-O-Mg-Si-vs-Ejecta}
\end{figure}

We constructed a single image by reprojecting the 14 observations to a common axis and merging the event files using the \textit{CIAO} script \textit{merge\_obs}. From this, we extracted narrow band images using the \textit{CIAO} tool \textit{dmcopy} with 1$''$ binning for each of the band energies listed in Table \ref{tab:BandEnergies}. 
In Figure \ref{fig:RGB-O-Mg-Si-vs-Ejecta} (left) we show spatial separation of O and Mg emission from the emission from Si. Distinct pockets of red, green, and blue are visible. The right panel shows the emission from S, Ar, and Ca where no clear spatial separation is apparent. This may point to the intermediate mass elements originating from bordering burning layers leading to similar distributions of ejected material. The emission from the ejecta elements follows a different distribution from the strongest O and Mg emission. We note that the O band might contain some emission from Fe L-shell lines as well. {\it Chandra's} spectral resolution is not sufficient to disentangle O from Fe L-shell lines, but for the purposes of this paper, we simply label the region ``O."

\subsection{Equivalent Width Images}
We then constructed equivalent width images (EWIs) which highlight line emission relative to the strength of the continuum using the procedure outlined in \cite{Hwang2000} and \cite{Winkler2014}. From the merged event file we created narrow band images binned with 1$''$ pixels for each of the energy bands listed in Table \ref{tab:BandEnergies}. We estimate the continuum by linearly interpolating between the normalized low and high energy bands to calculate a predicted continuum image. We then subtracted this continuum image from the normalized line image and the excess was then divided by the predicted continuum, and smoothed the resulting image with a 2-pixel Gaussian.

\begin{figure}
    \centering
    \includegraphics[width=\columnwidth]{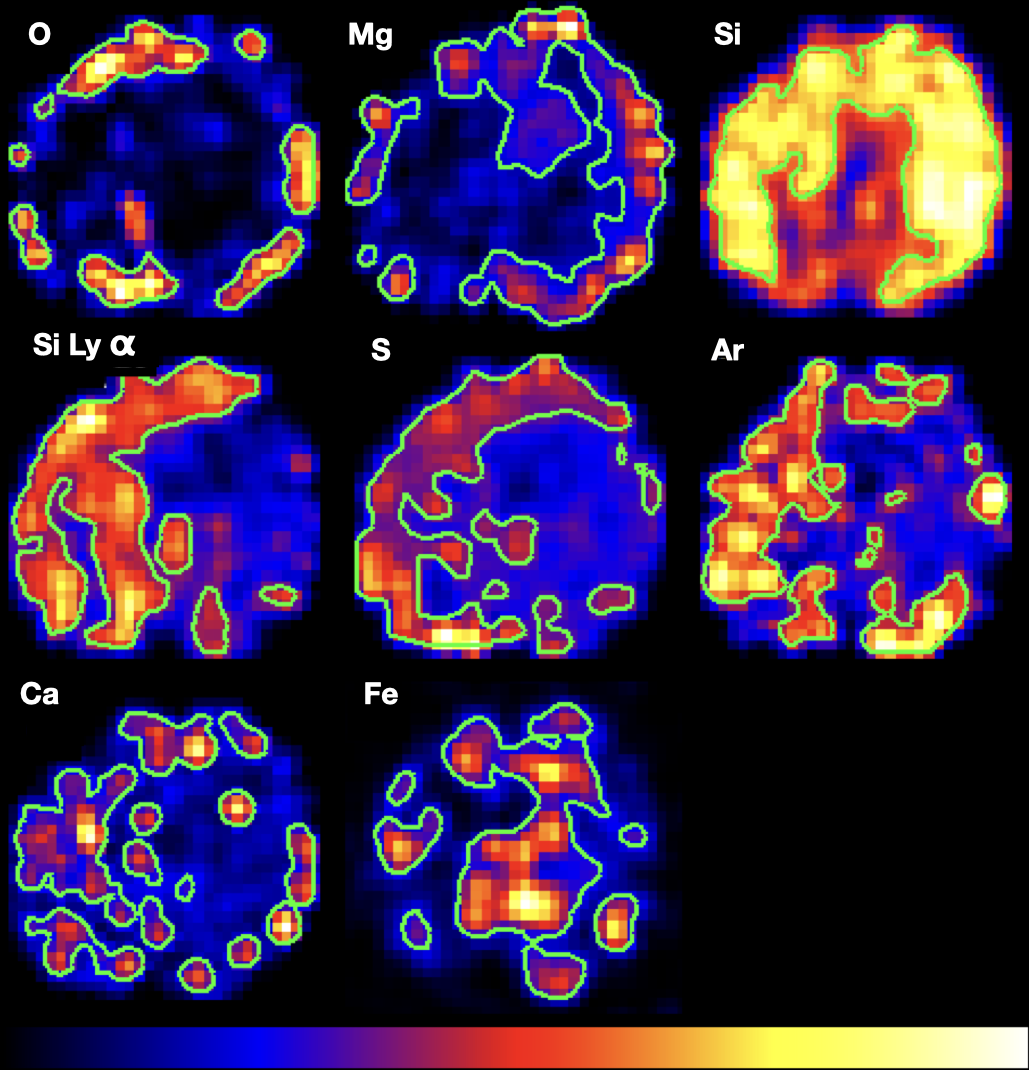}
    \caption{Equivalent width line images with the contours containing the highest intensity regions overlaid. See the text for details.}
    \label{fig:EW-2pix-ContourRegions}
\end{figure}

Figure \ref{fig:EW-2pix-ContourRegions} shows the resulting EW images. The O and Mg images show enhancements that appears as a ring tracing the edge of the remnant with minimal contributions from the center.
The Si image shows enhancement filling the remnant; however, it is most significant in a horseshoe morphology with lesser enhancement in the south extending to the center. The Si Ly$\alpha$ line is strong enough to create a similar image and shows stronger enhancement in the fainter eastern half. S, Ar, and Ca all show most significant enhancement on the eastern side with pockets filling the remnant. There is a notable hole in each image which coincides with the pocket of interior enhancement seen in the Mg image as the inward extension of the contour in the north. The Fe K$\alpha$ image shows prominent emission filling the remnant with the brightest emission located to the south west of the remnant center.

\begin{figure*}
\gridline{\fig{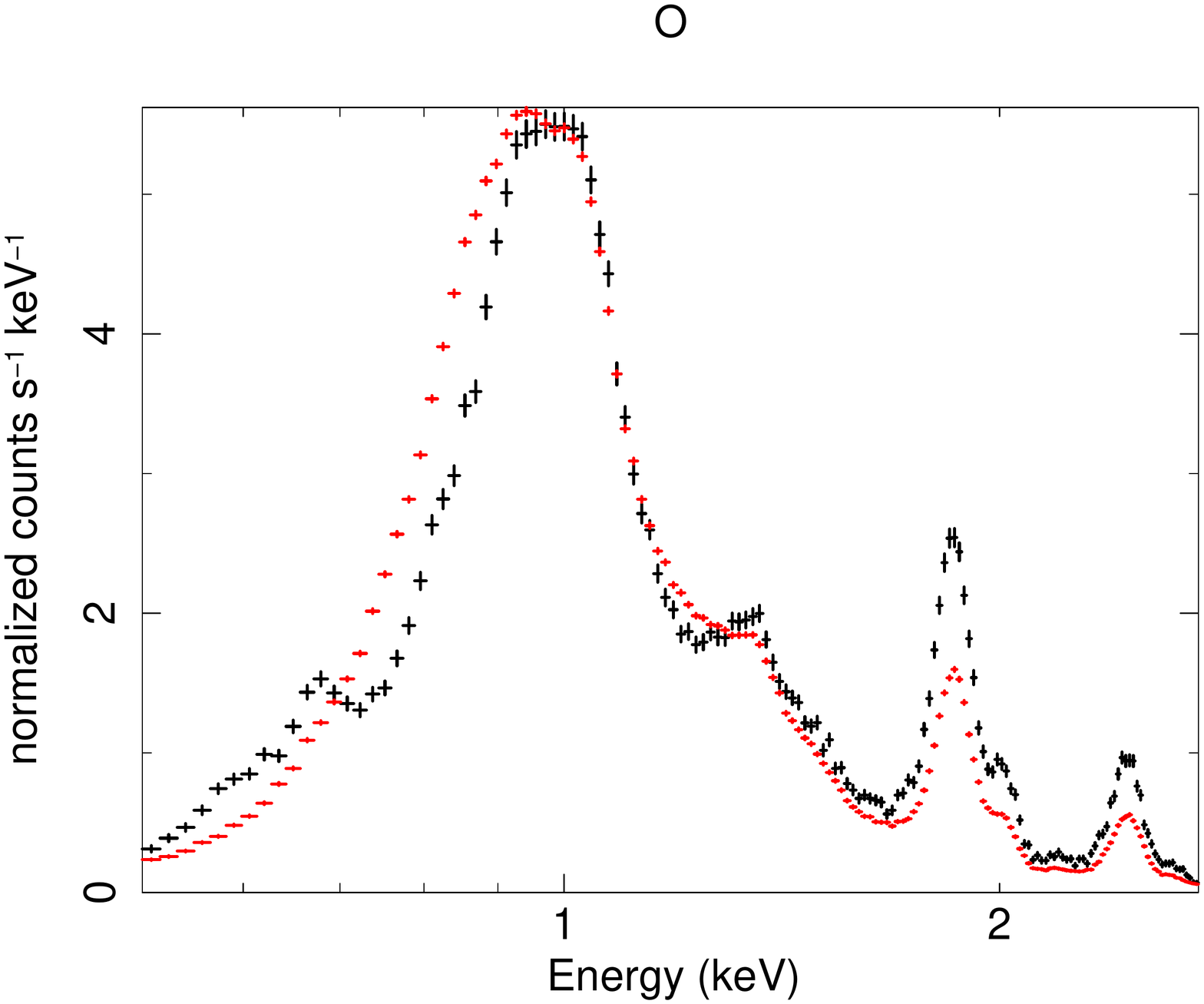}{0.49\textwidth}{(a)}
        \fig{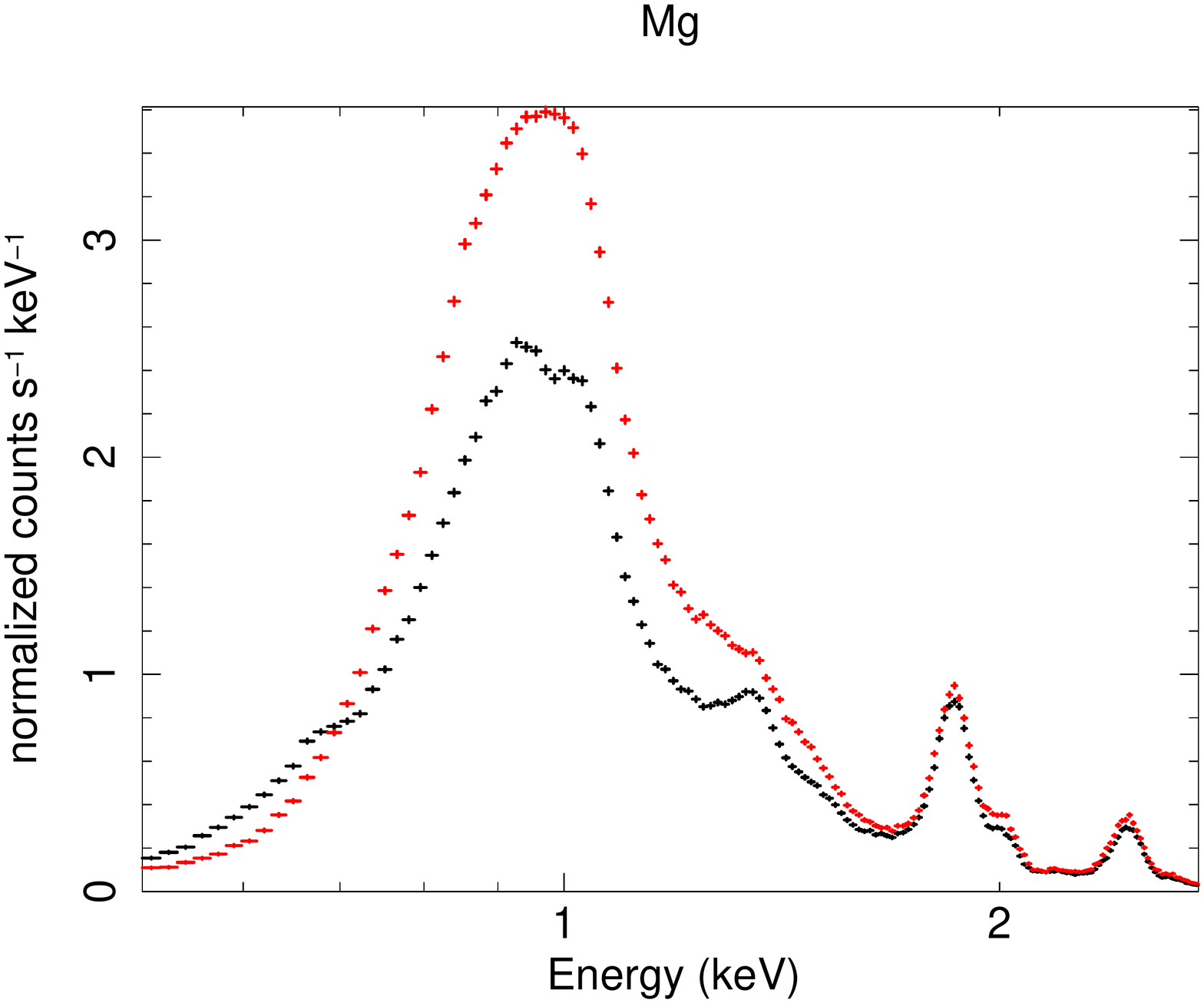}{0.49\textwidth}{(b)}
        }
        
        \gridline{\fig{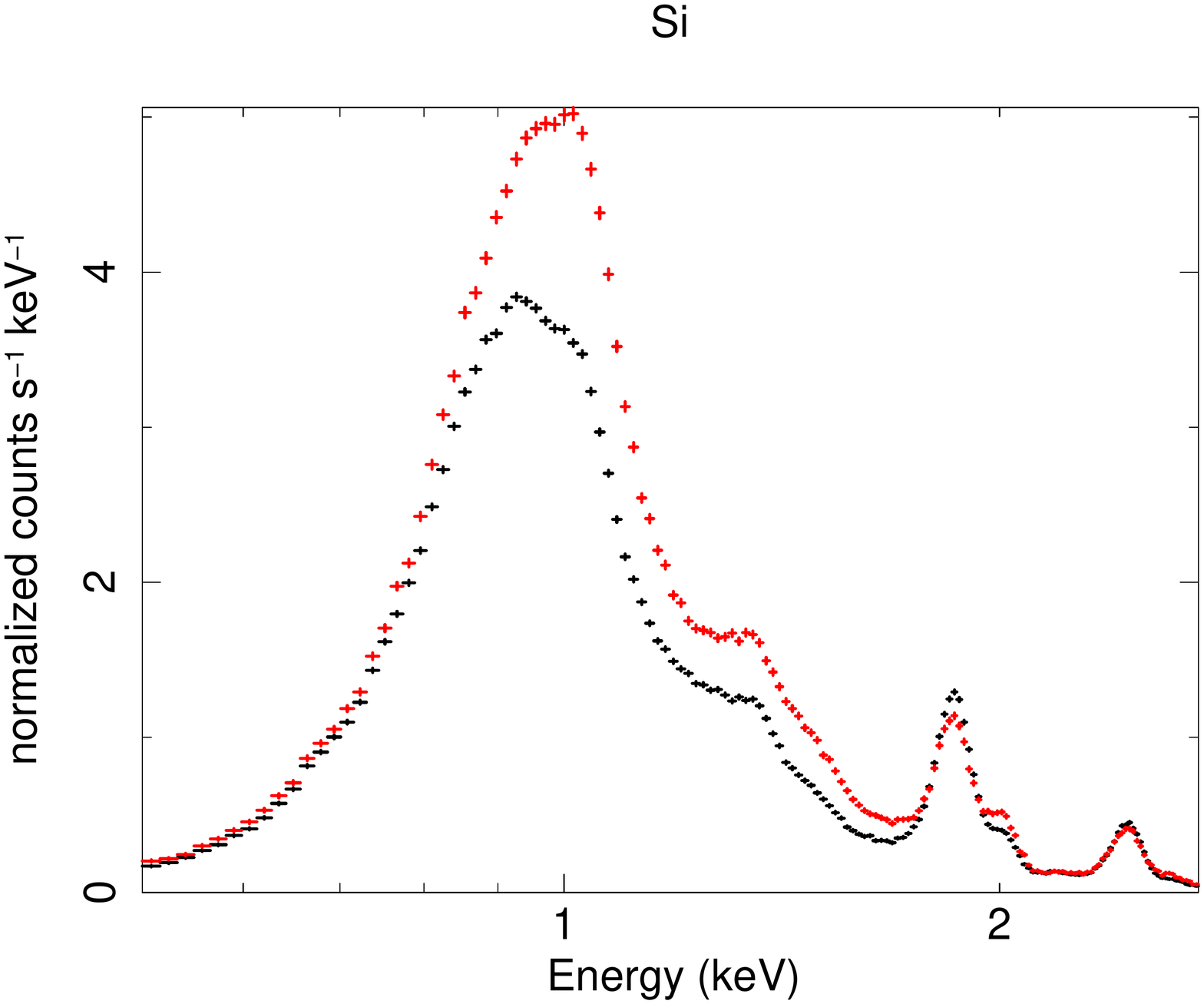}{0.49\textwidth}{(c)}
                \fig{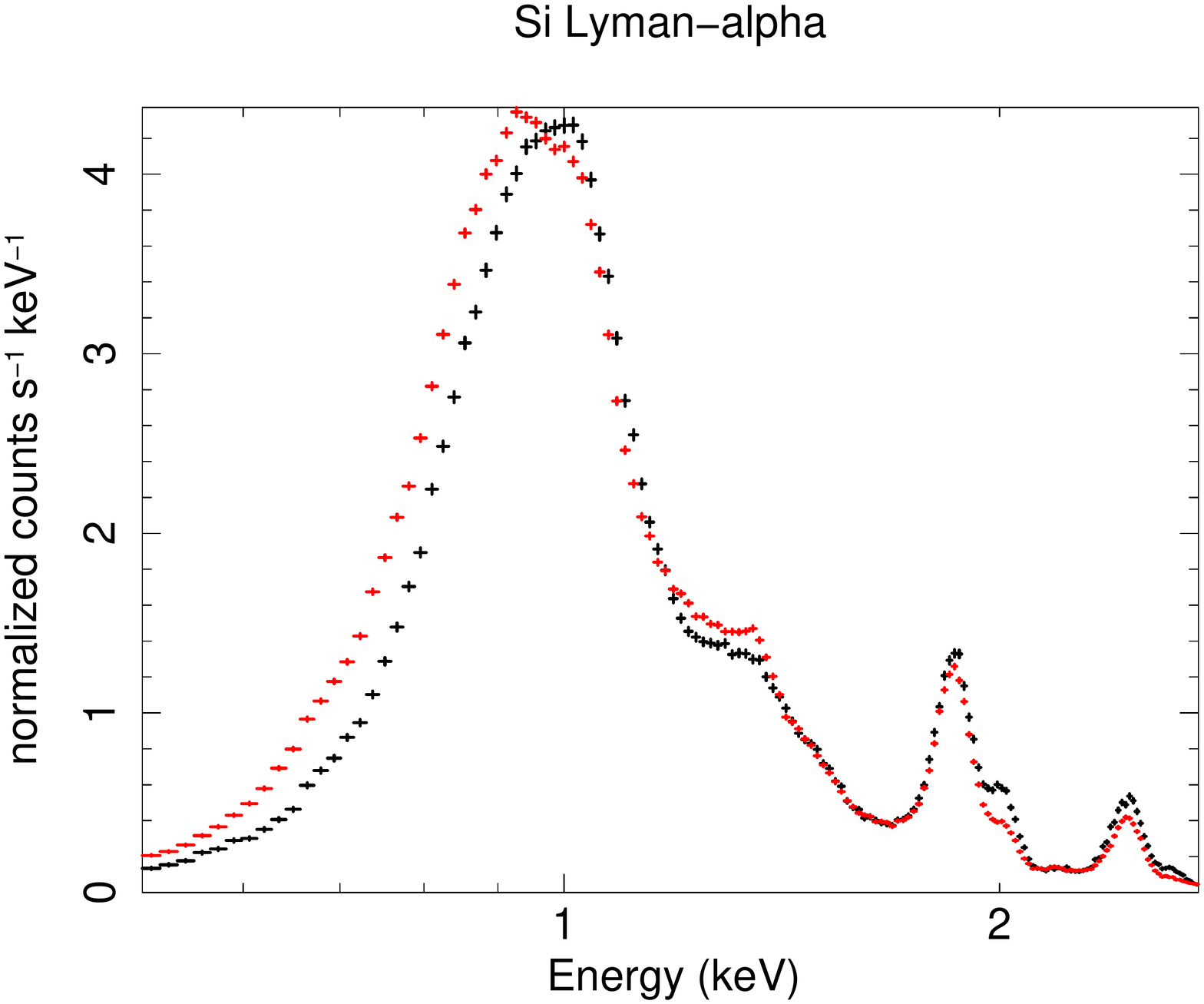}{0.49\textwidth}{(d)}
                }
        \gridline{\fig{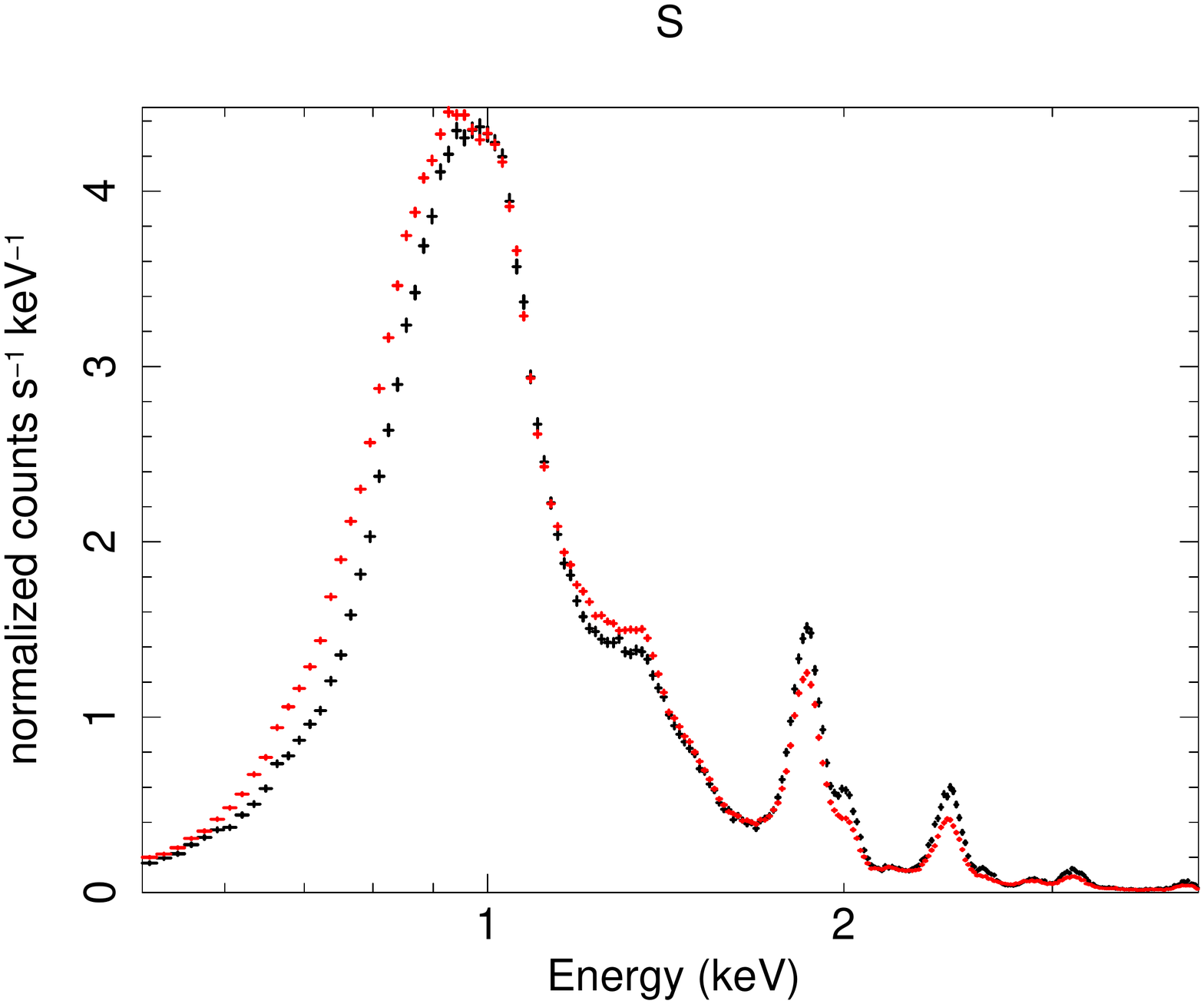}{0.49\textwidth}{(e)}
                \fig{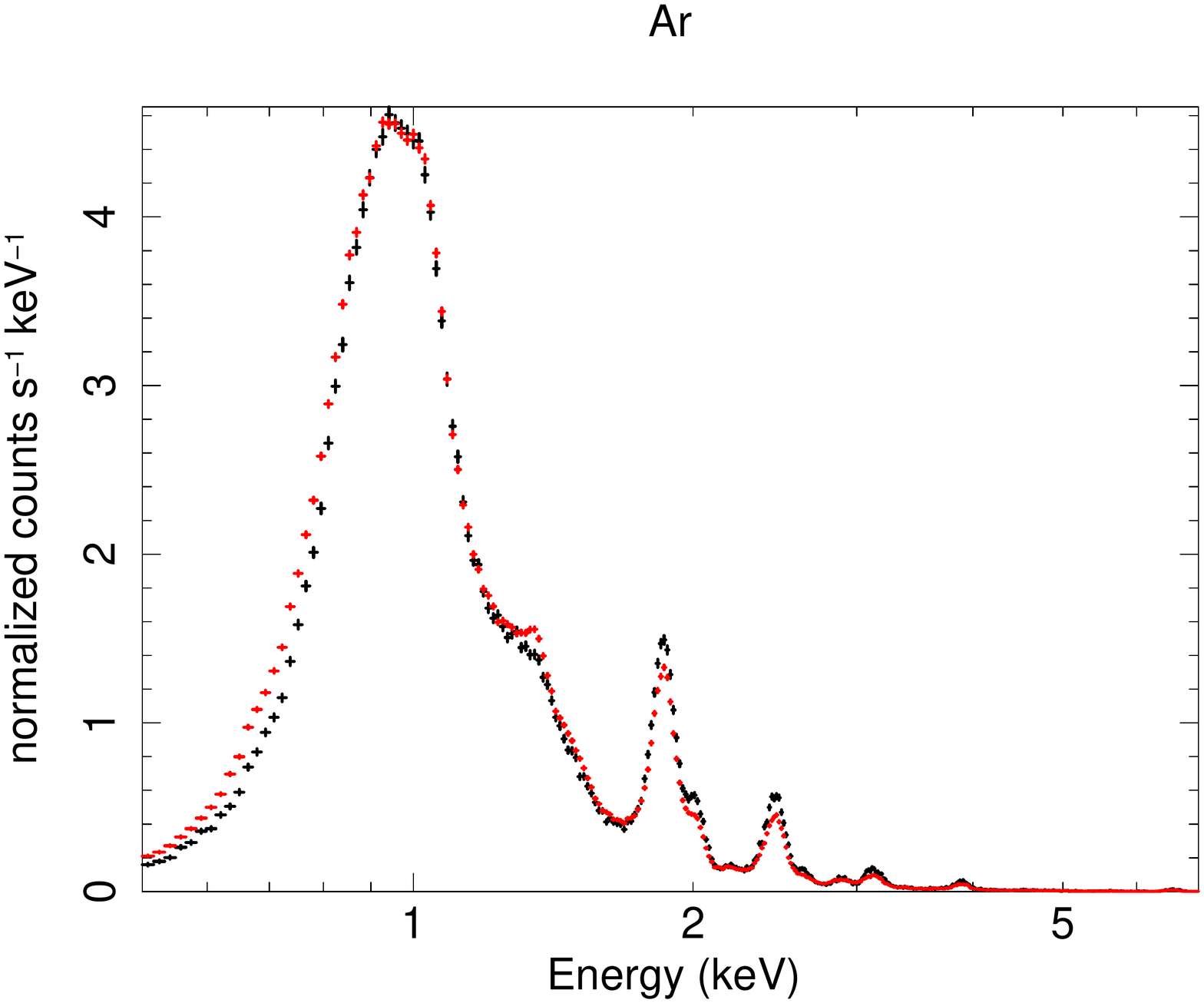}{0.49\textwidth}{(f)}
                }
    \caption{Comparison of spectra from the EWI contours (Figure \ref{fig:EW-2pix-ContourRegions}) shown in black and the rest of the remnant with these regions excluded shown in red. The spectra have been scaled with a multiplicative constant to align the 2.1-2.2 keV continuum in order to highlight the differences with the exception of the O Contours which have been scaled to align the 1keV peak in order to show the presence of the O and Mg lines which are weak or missing in the rest of the remnant.}
    \label{fig:ContourSpectra-subfigs}
\end{figure*}

\begin{figure*}
\gridline{\fig{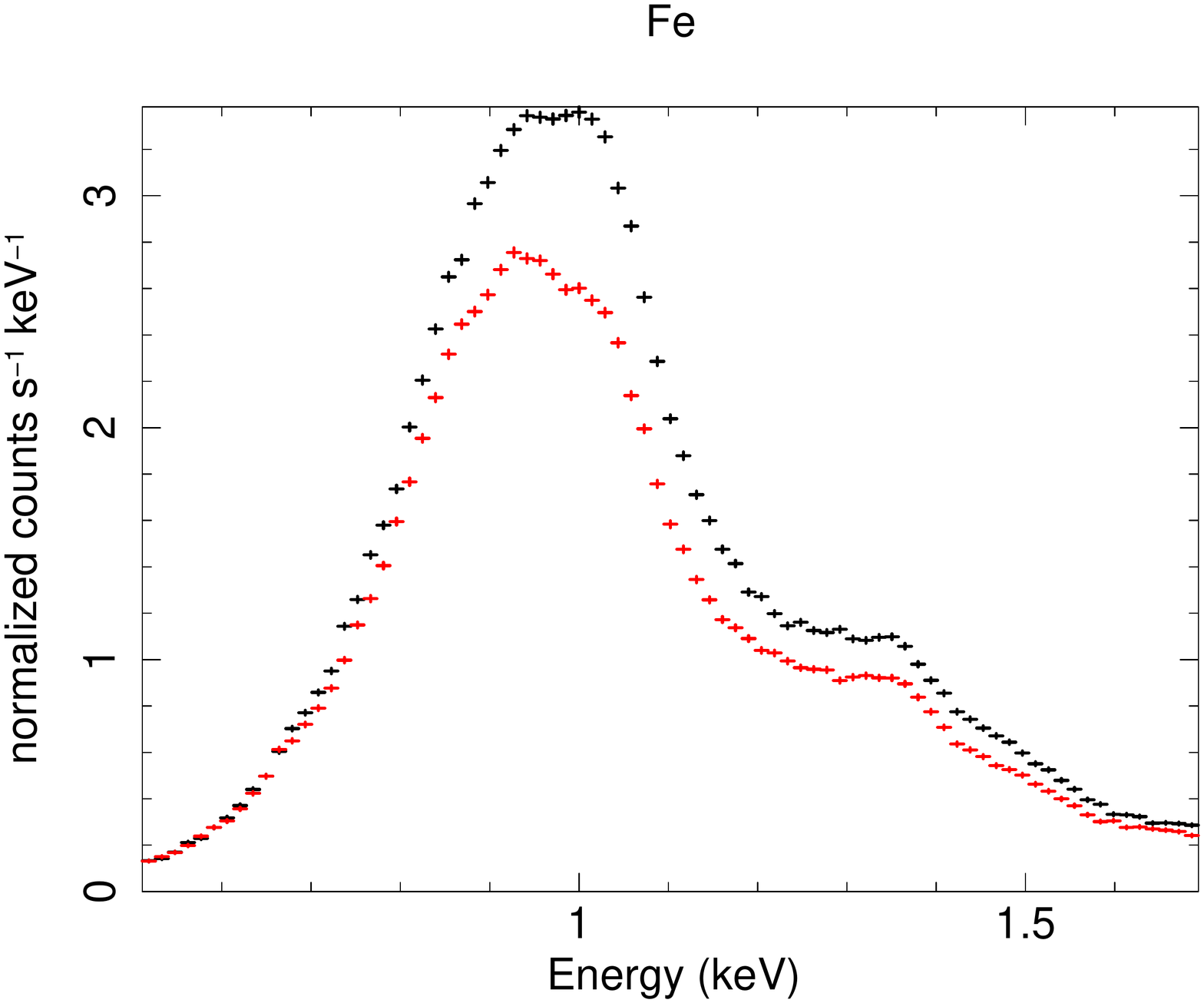}{0.49\textwidth}{(a)}
        \fig{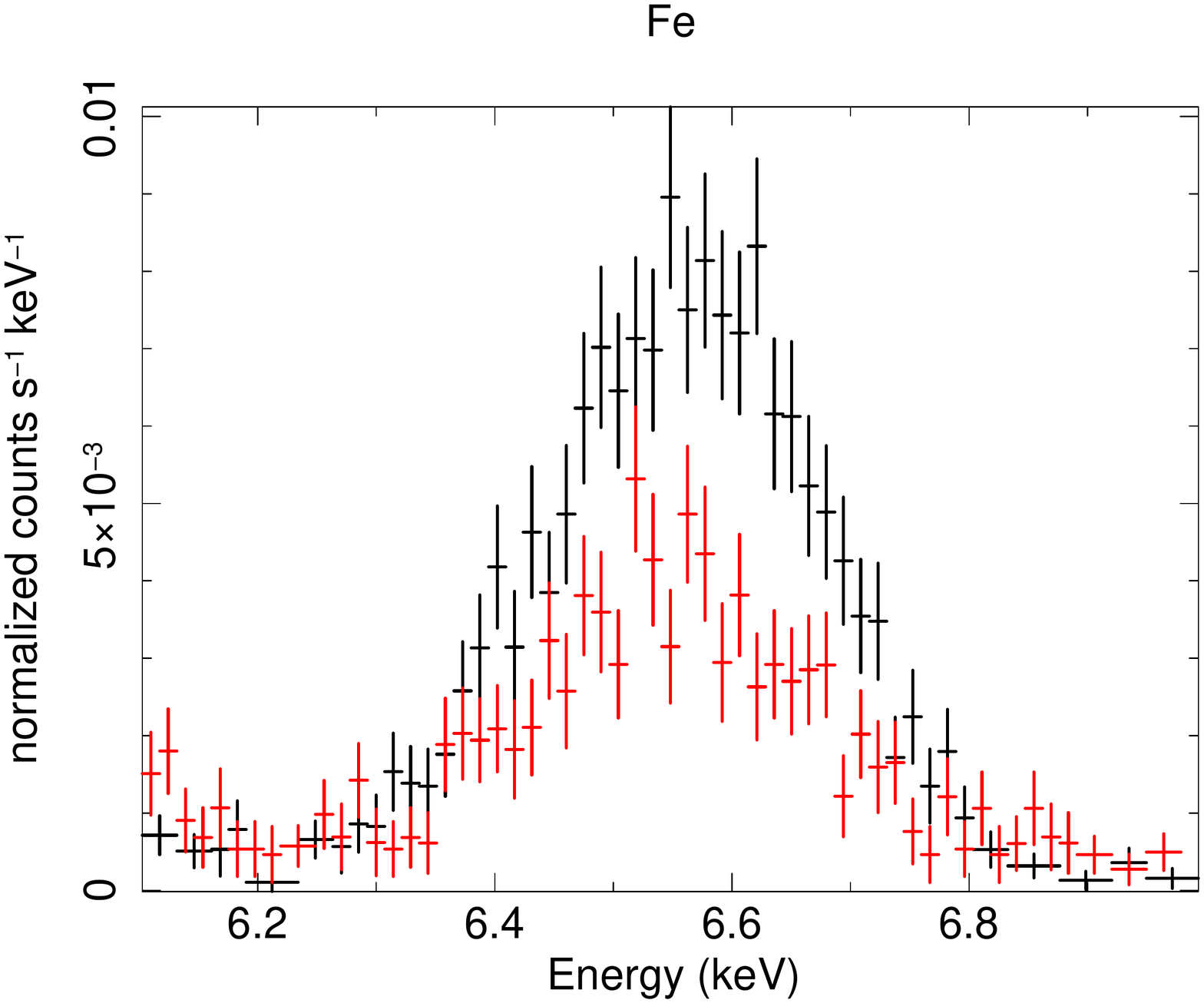}{0.49\textwidth}{(b)}
        }
    \caption{Comparison of the Fe K$\alpha$ rich contour regions from Figure \ref{fig:EW-2pix-ContourRegions} (black) and the rest of the remnant (red). The left and right panels highlight differences at low and high energies.}
    \label{fig:IronContourSpectra-subfigs}
\end{figure*}

To examine the correlation between each element we extracted spectra from the contours highlighting the enhanced regions shown in Figure \ref{fig:EW-2pix-ContourRegions} and compared to the spectra from a circle encompassing the full remnant with the contour regions excluded (Figures \ref{fig:ContourSpectra-subfigs} \& \ref{fig:IronContourSpectra-subfigs}).
To highlight the differences between the spectra rather than the overall brightness, scaling was changed where necessary in order to align the spectra at the level of the 2.1--2.2~keV continuum.

Primary differences occur in the soft $\sim 0.5 - 1.5$ keV range. The regions which show enhanced O lines correlate with the enhanced Mg regions such that each show prominent lines of the other. The heavier intermediate mass elements - Si through Ca - do not show any obvious correlation with enhancement of other elements, nor does the Fe L-shell or K$\alpha$ emission.

\begin{figure}
    \centering
    \includegraphics[width=\columnwidth]{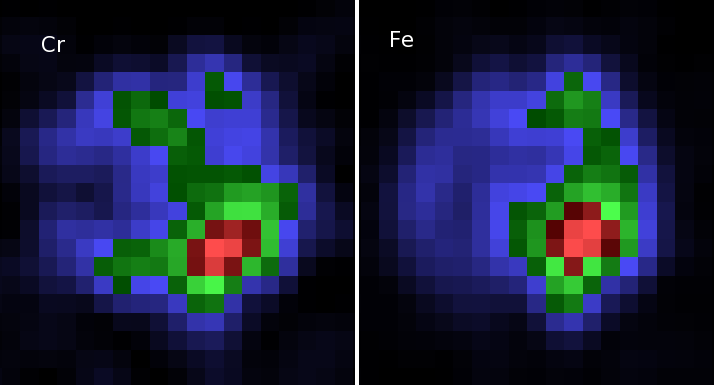}
    \caption{Cr (left) and Fe K$\alpha$ (right) narrow band line images. The pixel size is 2$''$ and the images have been smoothed with a 2 pixel Gaussian. The color scale is intensity.}
    \label{fig:Cr-Iron}
\end{figure}

We attempted to create an EWI for Cr ($\sim 5.6$ keV), but the line emission is insufficient to create an image with meaningful statistics given the minimal apparent line enhancement over the dim continuum emission at 5 keV. However, comparing images filtered by the line energies of Cr and Fe from Table \ref{tab:BandEnergies} (Figure \ref{fig:Cr-Iron}) we find a remarkable spatial agreement between the two, with both displaying prominent emission from a bright knot in the south west interior.

\subsection{Clustering}
\begin{figure}
    \centering
    \includegraphics[width=\textwidth]{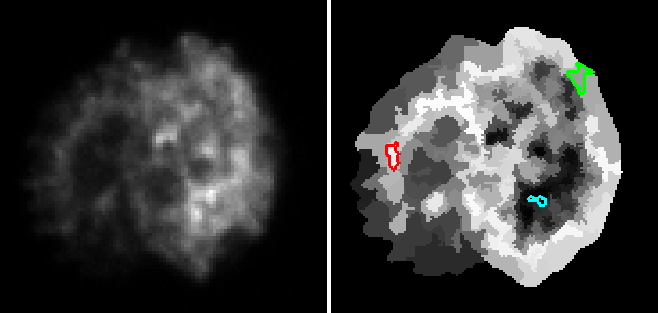}
    \caption{Broadband \textit{Chandra} input image (left) and the \textit{Contbin} generated region map (right). The grayscale in the region map indicates only region number and is related to the surface brightness and the order in which the regions were created by the binning algorithm. Spectra from the three small colored regions are displayed in Figure \ref{fig:ClusterRegionSpectraSample}.}
    \label{fig:ContbinMap}
\end{figure}

A method to identify and spatially separate emission from different origins was developed by \cite{Burkey2013}. Their work focused on separating emission from shocked CSM from that arising from ejecta. We build on this method and introduce the adaptive binning algorithm \textit{Contbin}  (\cite{Sanders2006}, \footnote{\url{https://github.com/jeremysanders/contbin/}}). This routine takes an input image and starting from the brightest pixel builds up a region following the surface brightness until the number of counts it contains meets a specified value. The result is a set of puzzle-piece like regions which cover an image while all containing a specified minimum number of counts. The input image we used was a reprojected and merged collection of all 14 \textit{Chandra} observations binned to 0.25$''$ pixels and spanning 0.3-10 keV. The \textit{Contbin} routine was then run on this image to create regions containing $\sim 7000$ photon counts. This produced over 200 regions (Figure \ref{fig:ContbinMap}).

For each region, we summed the number of photon counts belonging to narrow band energies (Table \ref{tab:BandEnergies}). The selected bands are O, Mg, Si, S, and the $\sim 1$ keV Fe complex. We normalized the number of counts in each band by the band energy width, and divided by a normalized associated continuum band to remove inherent bias due to continuum brightness. These continuum bands are the O low, Mg high, Si high, and S high bands. The Si and S lines counts were summed to produce 1 value, and the Fe complex line used the Si and S high bands summed for its continuum ratio. For each region we calculated a 4-D vector containing the O, Mg, Fe peak, and Si plus S ratios. Further processing was done by normalizing each vector component relative to the mean. We applied the \textit{kmeans} clustering algorithm (\cite{Lloyd1982}) to the set of vectors for a set number of clusters from 2-10. The number of clusters represents the number of components we wish to identify. The algorithm attempts to minimize the sum of the the 4D distance from the cluster center to each region associated with that cluster.

\begin{figure}
    \centering
    \includegraphics[width=\columnwidth]{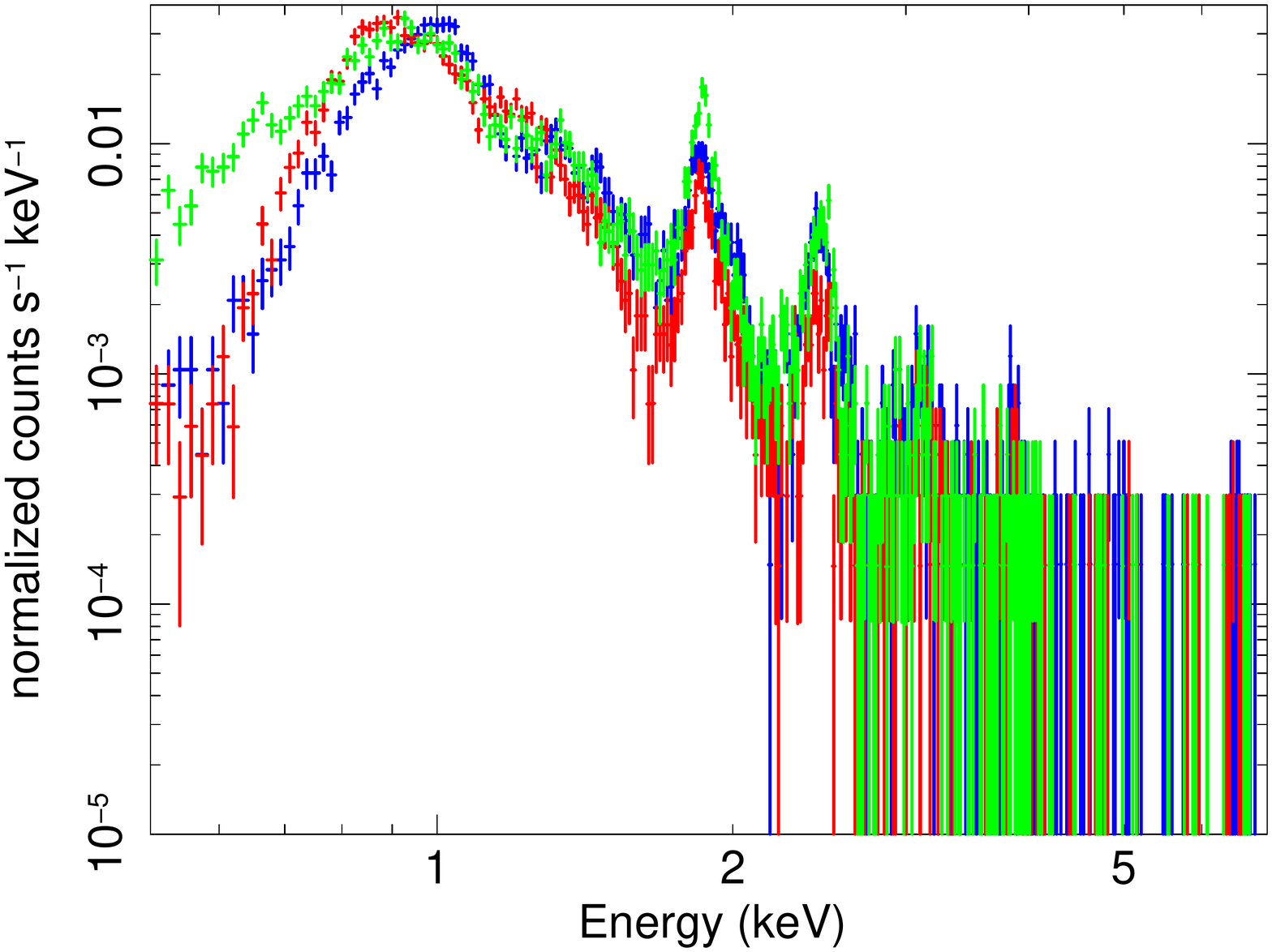}
    \caption{Sample spectra from 3 individual regions. Blue: bright south west knot segment (cyan region in Figure \ref{fig:ContbinMap}), Red: eastern arc segment, Green: Western limb segment. The regions are shown in Figure \ref{fig:ContbinMap}. The specific regions are arbitrary, however the selected sample show characteristic differences between spectra of the different clusters such as the low energy excess visible in the green spectrum and the energy of the peak intensity. Each spectrum contains over 7000 spectral counts and the binning shows how the signal to noise decreases at energies higher than the S band ($>2.8$ keV), yet allows for comparison between the O, Si, S, Mg, and 1 keV bands.}
    \label{fig:ClusterRegionSpectraSample}
\end{figure}

\begin{figure}
    \centering
    \includegraphics[width=\columnwidth]{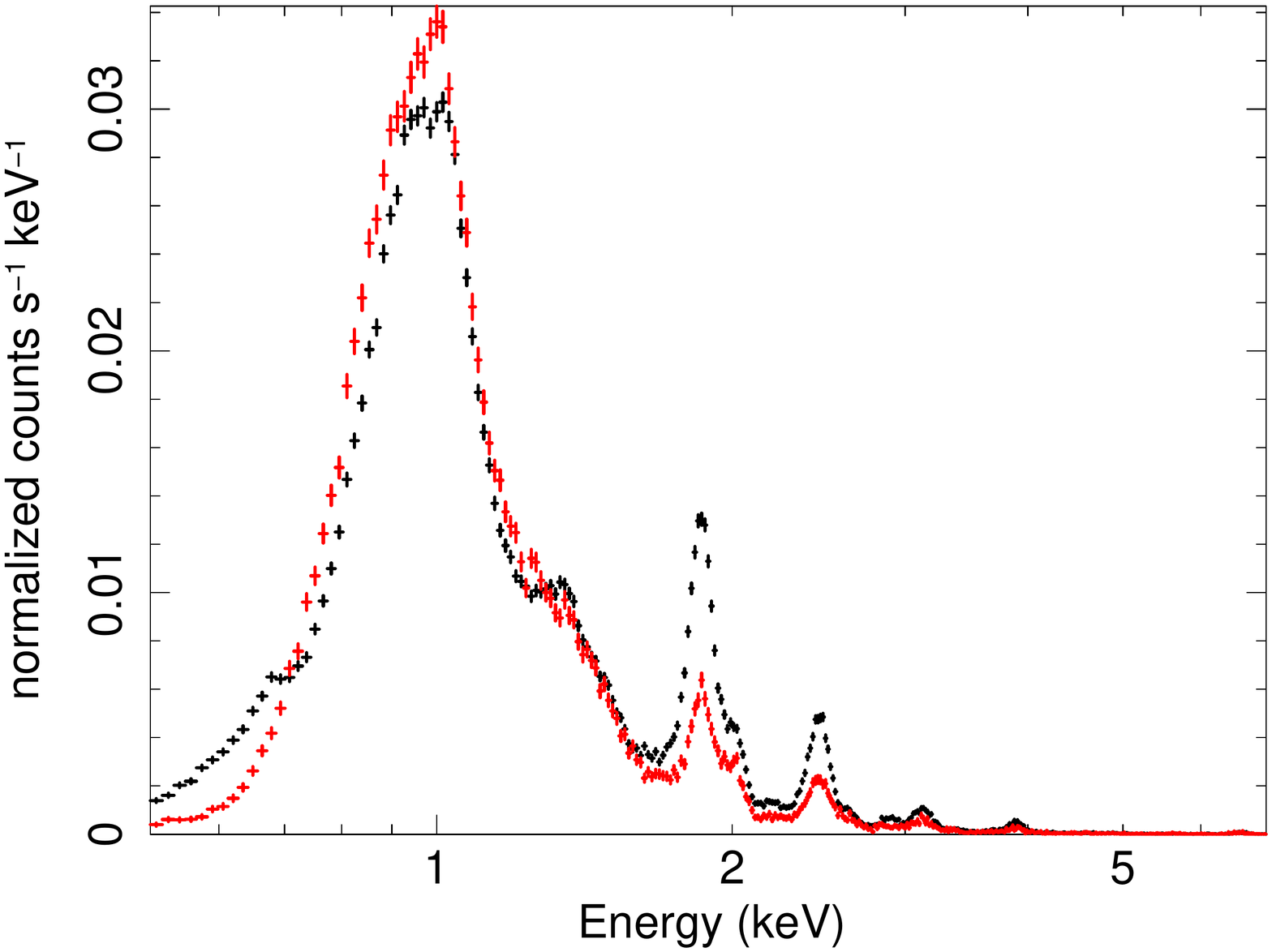}
    \caption{Comparing the merged spectra from 2 different clusters derived from the 10 cluster trial. Note the absence of O and Mg features in the red spectrum.}
    \label{fig:ClusterSpectraSample}
\end{figure}

\begin{figure}
    \centering
    \includegraphics[width=\columnwidth]{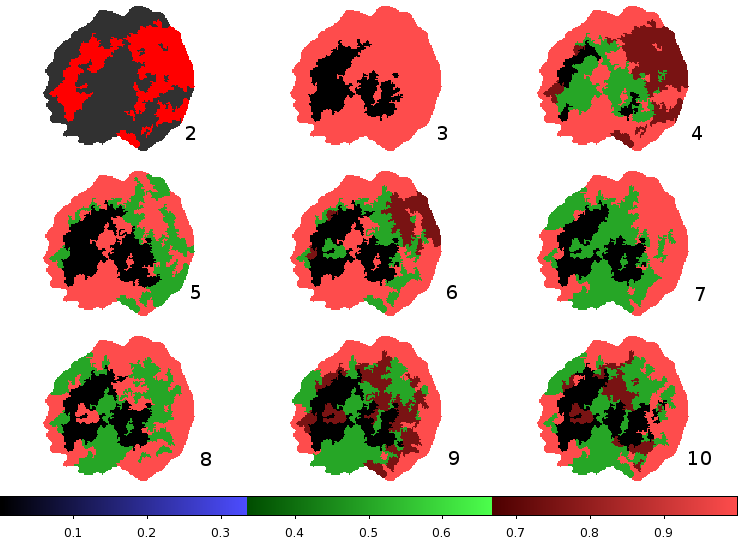}
    \caption{Visualization of the presence of O and Mg features in the merged spectra from each of the clustering runs. For display purposes a spectrum is valued 1 when both O and Mg features are clearly visible, 0.75 when one of the two is visible and the other is weakly visible, 0.5 when only one of the two features is visible, and 0 when neither are present. The 2 cluster tile is colored differently since both clusters show similar O and Mg features. When larger number of clusters are used, allowing for more subtle differences in the spectra to become important the O and Mg features are found to be most visible on the outer edge of the remnant.}
    \label{fig:ClusterClassifications}
\end{figure}

We extracted spectra from each region with background taken from an annulus surrounding the remnant. These individual regions display varying characteristics (Figure \ref{fig:ClusterRegionSpectraSample}). We merged the spectra from regions identified as belonging to a common cluster using the CIAO routine \textit{combine\_spectra}, and examined by eye the combined spectra (Figure \ref{fig:ClusterSpectraSample}) to search for the presence of O and Mg line features. We value a cluster at 1 if both O and Mg features are seen, 0.75 if 1 is seen and it is unclear whether the other is present, 0.5 if only one is present, and 0 if neither feature is visible. Results are displayed in Figure \ref{fig:ClusterClassifications}.

\begin{figure}
    \centering
    \includegraphics[width=\columnwidth]{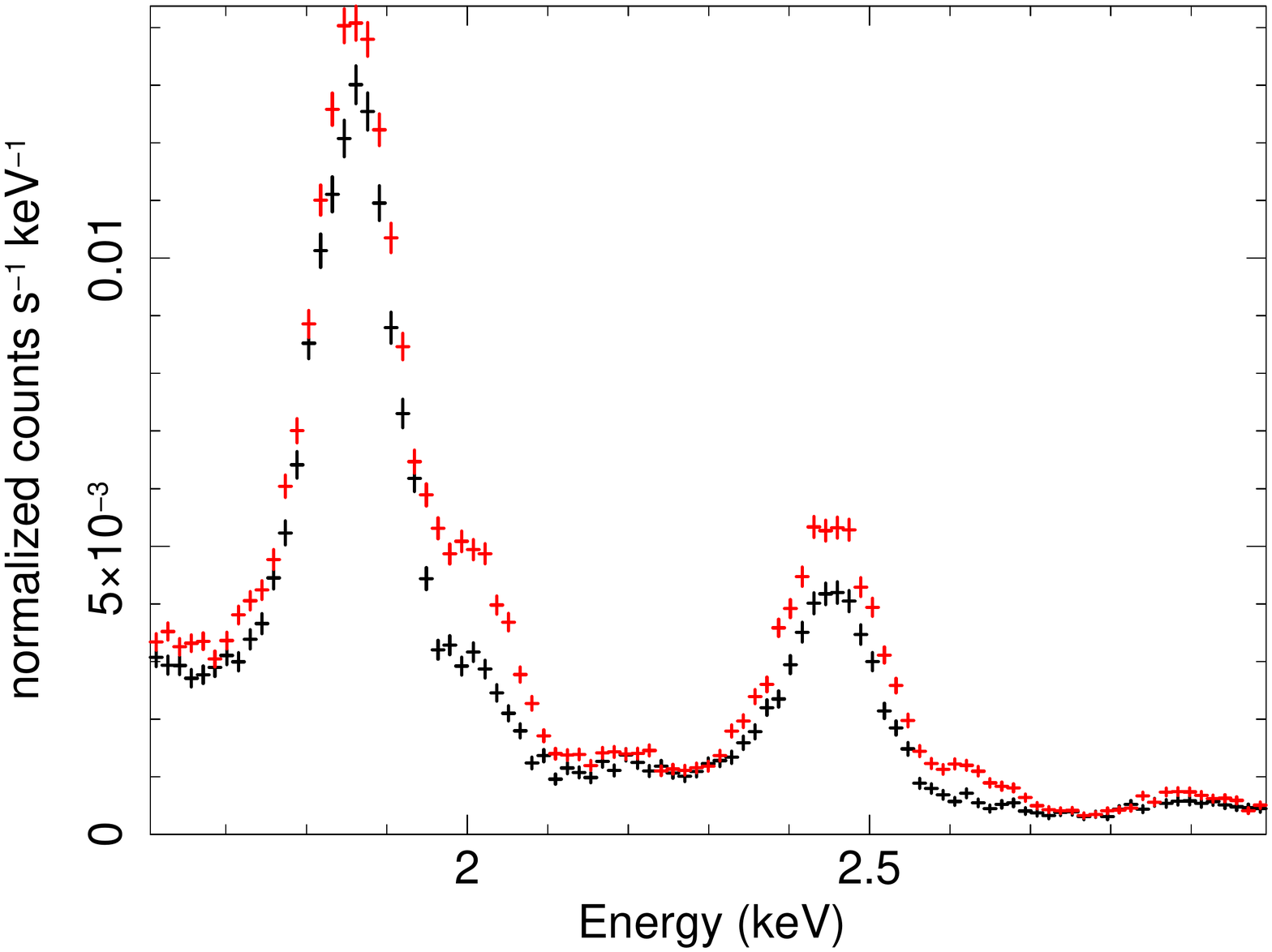}
    \caption{Sample of the merged spectra from 2 of the clusters derived from the 9 cluster run. There is a notable difference in the strength of the Si Ly$\alpha$ line at 2.01 keV.}
    \label{fig:LineWingStrength}
\end{figure}

In agreement with the EWIs, the regions bordering the edge of the remnant
show prominent O and Mg line features while the interior regions, particularly for cluster numbers greater than 6, show either Mg and no O or neither of the two features. An additional result was the identification of clusters which contained strong line wings belonging to additional Si and S transitions, such as the Si Ly$\alpha$ line at 2.01 keV. See Figure \ref{fig:LineWingStrength} for an example.

We find that for larger numbers of clusters, spectra continue to separate into groups with similar characteristics. With 7 or more clusters, the O and Mg features are found in spectra only derived from regions bordering the edge of the remnant, with the most prominent features found on the western side. The strongest Ly$\alpha$ lines of Si and S are found from a band which runs from the bottom of the remnant towards the center.

\begin{figure}
    \centering
    \includegraphics[width=\columnwidth]{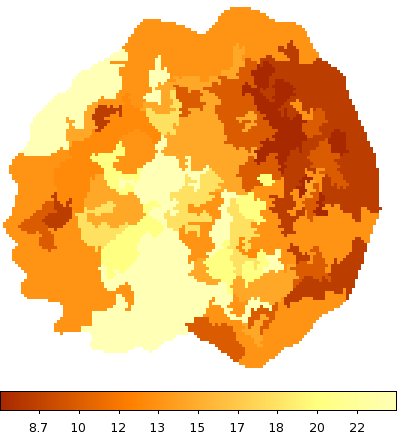}
    \caption{Map of the ionization timescale in units of $10^{10}$s cm$^{-3}$. The individual input spectra correspond to the groups identified as similar by the \textit{kmeans} algorithm for the 10 clusters run. See the text for a description of the model.}
    \label{fig:10Clusters-Ionization}
\end{figure}

\begin{figure*}
\gridline{\fig{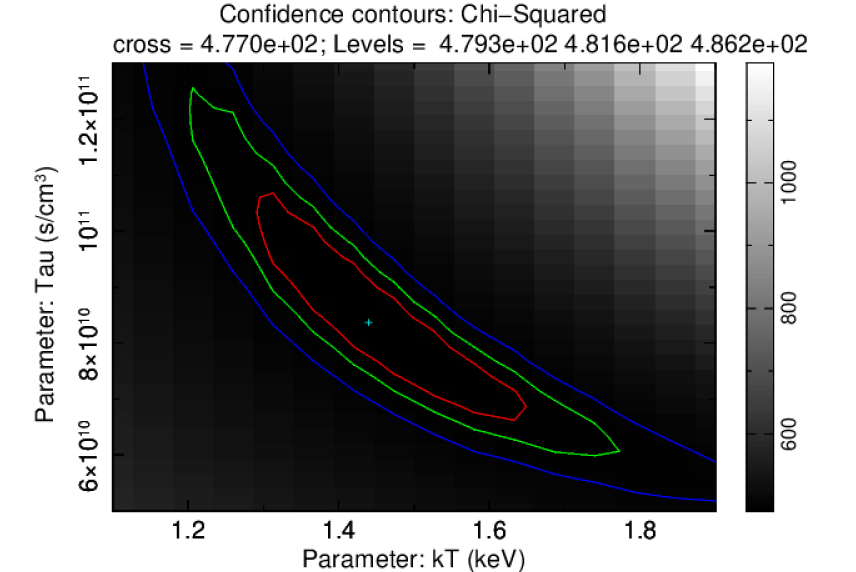}{0.49\textwidth}{(a)}
        \fig{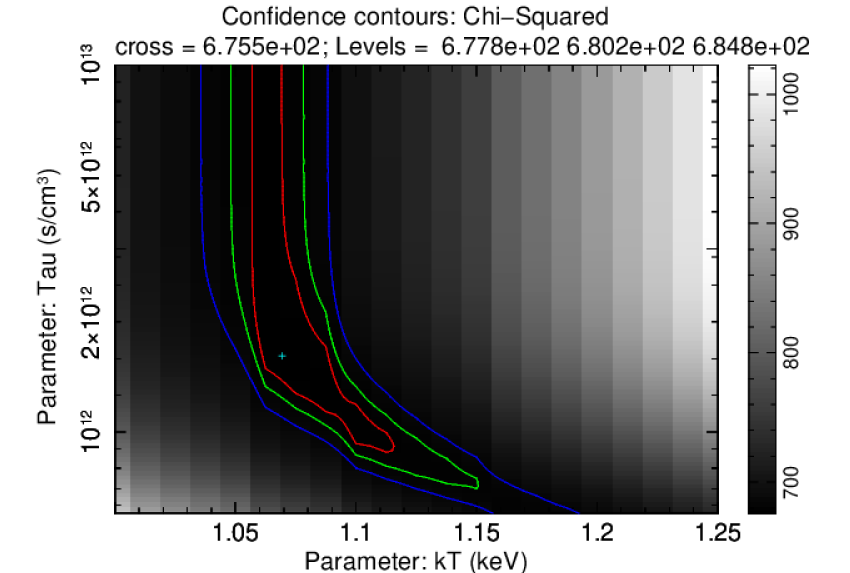}{0.49\textwidth}{(b)}
        }
    \caption{Error contours for the two clusters which found the smallest (left) and largest (right) ionization timescale from Figure \ref{fig:10Clusters-Ionization}.}
    \label{fig:Ionization-contours}
\end{figure*}

We investigated these regions in more detail by fitting each of the merged 10 clusters spectra with a VNEI model over the limited energy range $1.7-2.8$ keV, a range that covers the Si and S lines. \cite{Williams2017} showed that with strong emission lines from Si and S, this range is sufficient to constrain the ionization state of the plasma. The ionization timescale is mapped in Figure \ref{fig:10Clusters-Ionization}. The ionization timescale correlates inversely with brightness, with the bright western half of the remnant displaying a smaller timescale than the fainter eastern half. This is counter intuitive since the ionization timescale should correlate with electron density and brightness. We investigated the constraints on the ionization fits with the error contour plots in Figure \ref{fig:Ionization-contours}. The ionization timescale is not tightly constrained for the fainter central region, yet it favours a larger timescale and smaller temperature. The large difference in brightness between the eastern half of the remnant and the region we identify with the largest ionization timescale likely plays a role. The excess continuum emission unaccounted for in our 1-component model may effectively weaken the appearance of the Ly$\alpha$ line leading to a systematically lower timescale in the brighter regions. Despite these uncertainties, this large ionization timescale region is spatially coincident with a similar region identified as an inner ring by the component analysis of \cite{Yamaguchi21} supporting the presence of substructure within the remnant.

\section{Spectral modeling}\label{sec:SpectralModeling}
The EWIs presented in Figure \ref{fig:EW-2pix-ContourRegions} highlight regions where the line emission is the strongest relative to the continuum. The spectra extracted from contours surrounding the brightest line emitting regions shown in Figures \ref{fig:ContourSpectra-subfigs}-\ref{fig:IronContourSpectra-subfigs} shows that the EWIs represent real features in the underlying spectra. We cannot immediately attribute areas of line brightness revealed by the EWIs to locally enhanced abundances in these regions. It is possible that they may result from the effect of ionization timescale and plasma temperature in enhancing the emission.
To examine the underlying abundances in more detail, we fit the spectra with an absorbed multi-component non-equilibrium ionization plasma model \textit{VNEI} (\cite{Borkowski2001}). The absorption from the Milky Way is modeled by a {\it Tbabs} model with Wilms abundances (\cite{Wilms2000}) fixed to $6 \times 10^{20}\ \text{cm}^{-2}$ (\cite{Dickey1990}) and a free to vary LMC component modeled with {\it Tbvarabs} with the LMC average abundances from \cite{Dopita2019}. Fixing one {\it vnei} component with the average LMC abundances, a second component composed of only the intermediate mass elements Si, S, Ar, and Ca present and allowed to vary, and a 3rd Fe only component. We process the observations individually, creating response files and background for each. 

\begin{figure}
    \centering
    \includegraphics[width=\columnwidth]{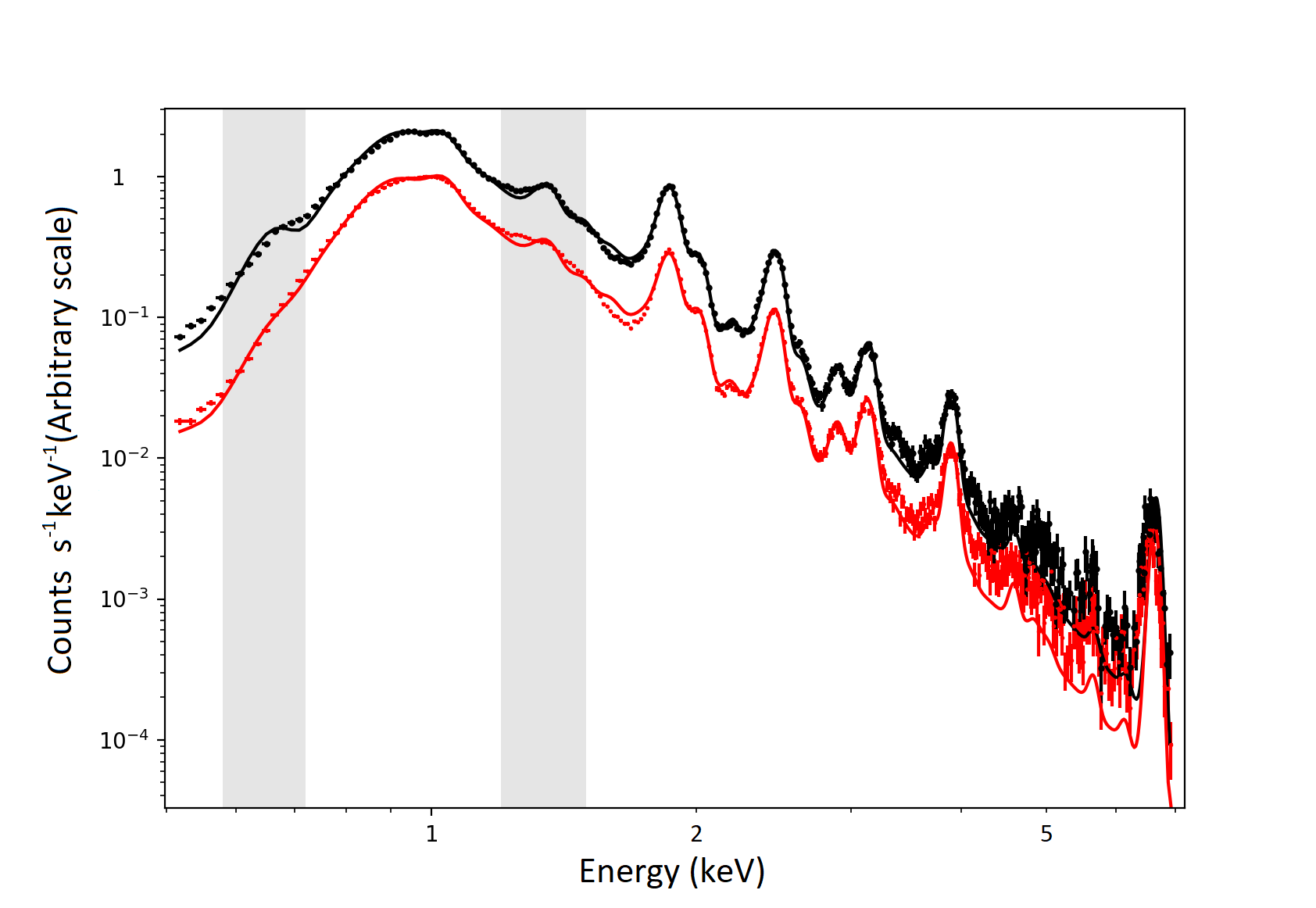}
    \caption{Spectral models from Table \ref{tab:SpectralFit}. The black spectrum is from the Mg contours and the red spectrum is from the rest of the remnant. The spectra have been merged for display purposes and the red spectrum has scaled down by a factor of 3 to avoid overlap. The model derived from simultaneously fitting the individual observations is overlaid on the merged spectrum. We have highlighted in grey the O and Mg bands which show prominence in the contour spectrum.}
    \label{fig:SpectralFits}
\end{figure}

Comparing the best fit models from the EWI contour spectra for Mg, we find enhanced O and Mg from the outer ring contours compared to the interior. The best-fit models from the Mg enhanced contours and the remnant with these regions excluded are listed in Table \ref{tab:SpectralFit}, and the spectra are shown in Figure \ref{fig:SpectralFits}.

\begin{table}[]
    \centering
    \begin{tabular}{c|c|c|c}
        Component	&	Parameter	&	Contours	&	Excluded\\\hline
Galactic	&	nH ($10^{22}$ cm$^{-2}$)	&	0.06 (Frozen)		&	0.06 (Frozen)		\\
LMC	&	nH($10^{22}$ cm$^{-2}$)	&	0.15	(0.147 - 0.153)	&	0.265	(0.26 - 0.27)	\\\hline
CSM	&	kT (keV)	&	0.9	(0.89 - 0.91)	&	0.834	(0.830 - 0.835)	\\
	&	O	&	1.9	(1.87 - 1.95)	&	0.63	(0.61 - 0.66)	\\
	&	Mg	&	0.69	(0.68 - 0.71)	&	0.42	(0.41 - 0.43)	\\
	&	Tau ($s$ cm$^{-3}$)	&	1.56E+11	(1.54E11 - 1.58E11)	&	1.84E+11	(1.81e11 - 1.88e11)	\\
	&	Norm (cm$^{-5}$)	&	5.54E-03	(5.51e-3 - 5.55e-3)	&	9.72E-03	(9.70e-3 - 9.74e-3)	\\\hline
Ejecta	&	kT (keV)	&	3.32	(3.30 - 3.39)	&	4.6	(4.5 - 5.0)	\\
	&	Si	&	2.58	(2.55 - 2.61)	&	2.07	(2.01 - 2.15)	\\
	&	S	&	2.43	(2.39 - 2.48)	&	1.99	(1.95 - 2.05)	\\
	&	Ar	&	2.74	(2.62 - 2.89)	&	2.05	(1.95 - 2.16)	\\
	&	Ca	&	4.5	(4.29 - 4.76)	&	3.43	(3.3 - 3.62)	\\
	&	Tau	($s$ cm$^{-3}$)&	5.32E+10	(5.24e10 - 5.40e10)	&	5.01E+10	(4.94e10 - 5.07e10)	\\
	&	redshift	&	-3.80E-03	(-3.395e-3 -  --3.8e-3)	&	-2.72E-03	(-2.77e-3 - -2.68e-3)	\\
	&	Norm (cm$^{-5}$)	&	2.40E-03	(2.39e-3 - 2.42e-3)	&	2.78E-03	(2.72e-3 - 2.83e-3)	\\\hline
Fe	&	kT (keV)	&	7.4	(6.8 - 8.1)	&	10.9	(10.7 - 11.4)	\\
	&	Tau	($s$ cm$^{-3}$)&	7.03E+10	(6.86e10 - 7.20e10)	&	5.66E+10	(5.59e10 - 5.69e10)	\\
	&	redshift (tied)	&	-3.80E-03		&	-2.72E-03		\\
	&	Norm (cm$^{-5}$)	&	1.26E-03	(1.23e-3 - 1.29e-3)	&	2.10E-03	(2.06e-3 - 2.14e-3)	\\\hline
	&	$\chi_{\nu}(\nu)$	&	1.787 (2718)		&	2.346 (2874)		\\
    \end{tabular}
    \caption{Fits to the spectra extracted from the Mg EWI contours and the entire remnant with the contours excluded. The spectra are plotted in Figure \ref{fig:SpectralFits}. See the text for details. }

    \label{tab:SpectralFit}
\end{table}

\section{Multiwavelength Observations}
\begin{figure}
    \centering
    \includegraphics[width=\columnwidth]{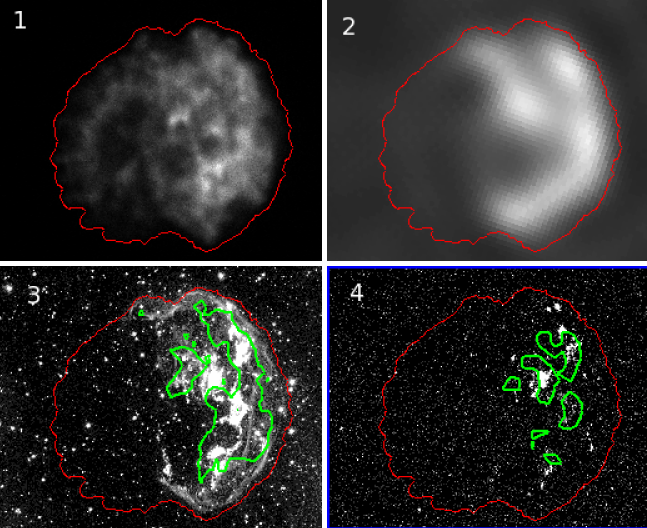}
    \caption{Multiwavelength images with corresponding \textit{Chandra} contours. 1: Chandra broadband image, 2: {\it Spitzer} 24$\mu$m deconvolved (\cite{Williams2014}), 3: {\it HST} H$\alpha$, 4: {\it HST} [O III]. The red contour is the {\it Chandra} broadband size, the green contours are the corresponding X-ray line contours. The H$\alpha$ green contour is the 2.1--2.3 keV continuum emission.}
    \label{fig:SpitzerAndHubble}
\end{figure}

The multiwavelength picture of N103B reveals a complex structure (Figure \ref{fig:SpitzerAndHubble}). At X-ray energies, the remnant shows a circular morphology with the western half appearing significantly brighter than the east. At infrared wavelengths, only the western half is visible, with the deconvolved {\it Spitzer} 24 $\mu$m image revealing a limb and knot structure (\cite{Williams2014}). This same structure is seen at high spatial resolution in narrow band {\it Hubble} images from \cite{Li2017} and \cite{Blair2020}, where it is clear the bright emission knots, visible in F502N [O~III], F673N [S~II], F164N [Fe~II], arise from radiative shocks being driven into density enhancements. The H$\alpha$ (F657N) emission shows both the bright clumpy structure and a series of faint, smooth filaments which align with the \textit{Spitzer} outer shell structure and arise from non-radiative shocks along the primary shock front. The presence of these faint non-radiative shocks is an indicator that the shock is encountering at least partially neutral gas at these positions. This argument is supported by \cite{Ghavamian2017} using the Wide Field Integral Spectrograph (WiFeS) on the 2.3 m telescope at the Siding Spring Observatory in Australia to identify an intermediate-width component in the H$\alpha$ spectrum with a width of 145~km~s$-1$. The overall picture is consistent with the \textit{Spitzer} conclusion that denser CSM or ISM is being encountered by the shock on the western side of the SNR. This is the side of the SNR which is brightest in X-rays as well.  

\begin{figure}
    \centering
    \includegraphics[width=\columnwidth]{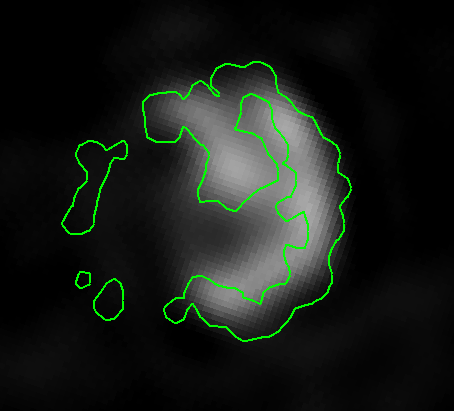}
    \caption{Contours from the Mg EWI overlaid on the deconvolved {\it Spitzer} 24 $\mu$m image.}
    \label{fig:Spitzer-MgEWI-Contours}
\end{figure}

\begin{figure}
    \centering
    \includegraphics[width=\columnwidth]{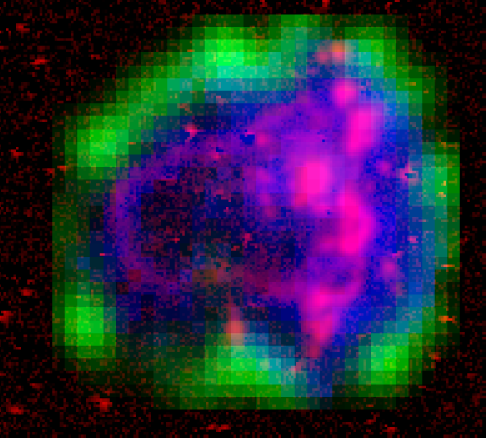}
    \caption{Image of N103B showing the reverse shock as traced by MUSE [Fe XIV] 5303 line emission (\cite{Seitenzahl19}) in red, and the \textit{Chandra} O EWI and narrowband line image in green and blue respectively.}
    \label{fig:ReverseShock}
\end{figure}

There is a clear offset between the distribution of O and Mg in the X-ray band which shows prominent emission tracing the blast wave, and the ejecta which fill the interior. The Mg EWI coincides with the deconvolved \textit{Spitzer} image (Figure \ref{fig:Spitzer-MgEWI-Contours}). Figure \ref{fig:ReverseShock} plots [Fe XIV] 5303 {\AA} emission from MUSE (\cite{Seitenzahl19}) in red outlining the location of the reverse shock in the remnant. The narrowband O line as seen by \textit{Chandra} is shown in blue, and the O EWI in green. The narrowband O line shows structure in the eastern half of the remnant overlapping the Fe reverse shock emission. However, this O line energy band may contain Fe L-shell emission lines.

\section{Discussion}
The integrated X-ray spectrum of N103B shows prominent lines from Si, S, Ar, Ca, and Fe. In addition, there are weaker lines from O and Mg. We have used spatial analysis using EWIs and the clustering technique described above to show that the O and Mg lines are found predominantly from a shell bordering the edge of the remnant while the intermediate mass elements fill the interior.

Despite the overwhelming difference in brightness of the remnant from west to east, the distribution of ejecta products does not reflect any obvious east-west asymmetries in the EWIs. Using multiplicative scaling to align spectra by their continuum emission brightness we have found strong agreement in the ejecta line strengths across the remnant. The Si Ly$\alpha$ EWI appears to favor the eastern half of the remnant. However, the density difference favoring the bright western half likely lessens the appearance of this line in the west by increasing the continuum level. The clustering analysis reveals a distinct separation in the location of the strongest O and Mg emission compared to the ejecta products in agreement with EWIs. We also find structure in the ionization timescale distribution. We find stronger Ly$\alpha$ lines characteristic of a larger ionization timescale from a north-south band in the eastern center of the remnant. 

This is in agreement with the results of \cite{Yamaguchi21} who used a component analysis technique to reveal similar structure. Where our results differ is our abundance measurement of the O and Mg rich emission. Allowing these parameters to vary, we find enhanced O and Mg from regions spatially coincident with the \textit{Spitzer} emission (\cite{Williams2014}), and the outer blast wave of the remnant. The uncertainties on the abundances are likely much larger than the values calculated from the data statistics in {\it XSpec}, and the spectrum shows residuals at low energies suggesting that these abundances should be taken rather as upper limits. However, the presence and coincidence of the strongest O and Mg line emission with the dust emission as seen in the 24 $\mu$m \textit{Spitzer} image, and the clear difference in spatial distribution compared to the ejecta products drives our argument that the remnant is encountering CSM, likely originating from mass loss from the progenitor system. 

The absence of N lines observable with \textit{Chandra} means we are unable to contrast the N to O ratios from the bright radiative knot studied in \cite{Blair2020}. Future high spectral resolution X-ray observations with \textit{XRISM} and \textit{ATHENA} will clarify the extent to which Fe L-shell lines are present in the ``O" band, and will allow a more accurate determination of the elemental abundances. Finally, IR observations with {\it JWST} at a spatial resolution comparable to that of {\it Chandra} will allow for a much more detailed morphological comparison between the emission from dust and gas in the remnant.

\pagebreak

\section{Conclusions}

Using both EWIs and a clustering technique we have found spatial differences in the distribution of O and Mg and the intermediate mass ejecta elements. We have confirmed that these regions are spectroscopically different as well. The O and Mg EWI and clustering images are morphologically similar to the {\it Spitzer} mid-IR emission from warm dust and the H$\alpha$ emission from non-radiative shocks, both of which arise from the interaction of the forward shock with the dense CSM, providing some evidence that the O and Mg seen in the X-rays also arise from a CSM origin.

The ejecta element EWIs show no preference for the bright western limb in terms of line enhancement, suggesting the difference in appearance is due to the enhanced density into which the remnant is expanding. The exception is the emission in the Cr and Fe K$\alpha$ lines, which are primarily concentrated in a knot of emission in the southwest. The O and Mg EWIs and spectra extracted from the most prominent regions contained therein present evidence of enhanced O and Mg relative to the average LMC abundances, suggesting the remnant is sweeping into CSM originating from mass loss from the pre-supernova system.

\begin{acknowledgments}
Support for this work was provided by NASA through Chandra General Observer Program Grant SAO G06-17064. B.G acknowledges the material is based upon work supported by NASA under award number 80GSFC21M0002. WPB acknowledges partial support from the JHU Center for Astrophysical Sciences during the time of this work. We thank the referee for their careful reading of the paper which improved its quality and clarity.
\end{acknowledgments}

\bibliography{References}{}
\bibliographystyle{aasjournal}



\end{document}